\documentclass[11pt]{article}
\usepackage[a4paper, margin=1in]{geometry}
\usepackage{iftex}
\ifPDFTeX\usepackage[T1]{fontenc}\fi
\ifPDFTeX\fi
\usepackage[usenames,dvipsnames,svgnames,x11names]{xcolor}
\usepackage{graphicx}
\usepackage{mathtools}
\usepackage{amssymb}
\usepackage{amsthm}
\usepackage{amsfonts}
\usepackage{gensymb}
\usepackage[caption=false]{subfig}
\usepackage{multirow}
\usepackage{booktabs}
\usepackage{array}
\newcolumntype{L}[1]{>{\raggedright\let\newline\\\arraybackslash\hspace{0pt}}m{#1}}
\newcolumntype{C}[1]{>{\centering\let\newline\\\arraybackslash\hspace{0pt}}m{#1}}
\newcolumntype{R}[1]{>{\raggedleft\let\newline\\\arraybackslash\hspace{0pt}}m{#1}}
\usepackage{xspace}
\usepackage{enumitem}
\usepackage{url}
\usepackage[colorlinks=true,linkcolor=blue,citecolor=blue,urlcolor=blue]{hyperref}

\newcommand{\ileq}[2]{#1~\refstepcounter{equation}\textup{(\theequation)}\label{#2}}

\newcommand{\Research}{\textsc{Research}}
\newcommand{\Assist}{\textsc{Assist}}
\newcommand{\Code}{\textsc{Code}}
\newcommand{\Chat}{\textsc{Chat}}
\newcommand{\AgenticTraceSim}{AgenticTraceSim}
\newcommand{\Spine}{\textsc{Spine}}
\newcommand{\Mid}{\textsc{Mid}}
\newcommand{\Scoped}{\textsc{Scoped}}
\newcommand{\Fanout}{\textsc{Fanout}}

\title{Understanding KV Residency in Production-Scale\\ Agentic LLM Workloads}
\author{
Mayank Arya\\{\normalsize Indian Institute of Science (IISc)}
\and Agrim Bari\\{\normalsize Microsoft M365 Research}
\and Anjaly Parayil\\{\normalsize Microsoft M365 Research}
\and Yogesh Simmhan\\{\normalsize Indian Institute of Science (IISc)}
\and Renee St.~Amant\\{\normalsize Microsoft M365 Research}
\and Victor R\"{u}hle\\{\normalsize Microsoft M365 Research}
\and Saravan Rajmohan\\{\normalsize Microsoft M365 Research}
}
\date{Email: mayankarya@iisc.ac.in; agrimbari@microsoft.com; aparayil@microsoft.com; simmhan@iisc.ac.in}

\begin{document}

\maketitle

\begin{abstract}
Agentic large language model (LLM) applications turn a single user request into a sequence of model and tool calls over a growing context. Retaining the key-value (KV) state computed for this context in GPU memory allows subsequent calls to reuse it rather than recompute it. However, retaining KV state across a serving fleet can require substantial GPU memory.
We characterize four production \textit{Microsoft M365 workload traces}, namely Deep Research, Enterprise Assistant, Coding Agent, and Enterprise Chat, spanning more than 121 million model calls. Agentic applications issue most calls within a session, and each call re-reads the growing context. In Deep Research, the application issues 89\% of model calls. Sessions have a median of 13 model calls, while Enterprise Chat, our lower-agency reference, has 333. The median context grows from approximately 10K tokens to 68K tokens by the $30^{th}$ call. About 80\% of context tokens are served from the prefix cache, and a predictable class of models accounts for 67\% of call-context KV.

This reusable KV state is large and remains idle for most of its lifetime. Since the production models are confidential, we use the public Llama-2-70B KV layout, with grouped-query attention and 16-bit values, to estimate the memory demand consistently across models. Under this assumption, the KV state of a single 99th-percentile Deep Research session exceeds the 80 GB high-bandwidth memory (HBM) capacity of an NVIDIA H100 GPU. Retaining the KV state of all live sessions at their peak, which we term the hold-all-live ceiling, requires an estimated 501.8\,TB of GPU HBM. However, model inference occupies only 5.6\% of the time for which this state is retained. Enterprise Chat has smaller sessions but approximately 349K concurrent live sessions, resulting in a similar capacity challenge.

Trace-driven replay shows that an idle cap of one hour reduces the hold-all-live ceiling by $3.6\times$, while increasing the fraction of calls that recompute their context from 4.07\% to 4.53\%. However, eviction alone is insufficient under tighter memory budgets. With a pooled budget equal to one tenth of the busiest hour's no-eviction demand, least recently used (LRU) eviction causes 63\% of Deep Research calls to recompute their context. Furthest-next-use (FNU), a clairvoyant policy based on Belady's rule, reduces this to 31\%. Multi-resolution retention (MRR) instead demotes the LRU session's KV state to an int4 representation within the same memory budget. MRR reduces the recomputation rate to 13.6\%, $2.3\times$ lower than FNU, despite selecting sessions using recency alone.

Placement imposes a separate limit on reuse. When the next call to the same model reaches a different deployment within the region, cached-context reuse drops from 89\% to 29\%. A Deep Research session accesses 4.5 deployments on average. Treating KV state as local to a deployment increases the first-touch floor, the fraction of calls that recompute their context even with unlimited memory, from 5.4\% to 20.2\%. Additional memory and better eviction cannot reduce this floor.

These findings motivate retention mechanisms that change the representation of KV state, together with deployment-aware routing or migration. Since the production traces cannot be shared, we plan to release \textit{AgenticTraceSim}, a parametric trace generator, and a residency replay framework for reproducible counterfactual studies.
\end{abstract}


\section{Introduction}
\label{sec:intro}

Production LLM serving is shifting from individual requests to persistent, tool-interleaved sessions. In conventional \textit{chat}, a user request opens a \emph{turn}: the serving system performs a short sequence of model and optional tool calls, and control returns to the user. In an \textit{agentic} workload, the request that opens a turn initiates a longer internal loop in which an orchestrator (e.g., one following the ReAct or Magentic pattern~\cite{react,magentic}) repeatedly invokes models and tools, appends their results, and decides when the task is complete~\cite{zhu2026tracelab,yuan2026workload,liu2026copilot}. A \emph{session} is a user's ongoing interaction with one application: a sequence of one or more turns that share a session identifier, together with the context they accumulate (Section~\ref{sec:fourtraces} defines calls, turns, and sessions precisely).

\begin{figure}[t]
  \centering
  \begin{minipage}{0.5\linewidth}
    \centering
\subfloat[Chat turn]{%
\includegraphics[width=0.7\linewidth]{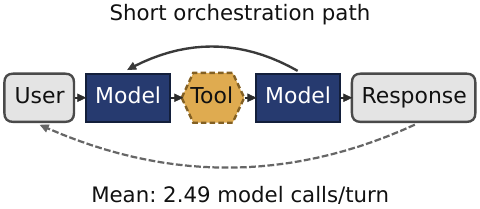}%
    \label{fig:autonomy-flow:chat}%
  }\\\vspace{-0.1in}
  \subfloat[Agentic turn]{%
    \includegraphics[width=1\linewidth]{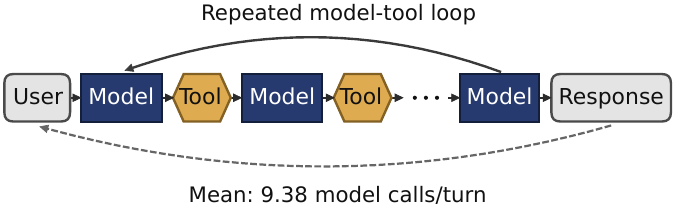}%
    \label{fig:autonomy-flow:agentic}%
  }
    \caption{Control flow within one user turn. A \emph{turn} runs from a user request until control returns to the user; a \emph{session} is the sequence of turns that share a session identifier, so this loop repeats once per turn. In-figure means are model calls per user turn (total model calls divided by total user turns), as in Table~\ref{tab:traces}.
    }
    \label{fig:autonomy-flow}
  \end{minipage}\quad
  \begin{minipage}{0.47\linewidth}
    \centering
    \includegraphics[width=1\linewidth]{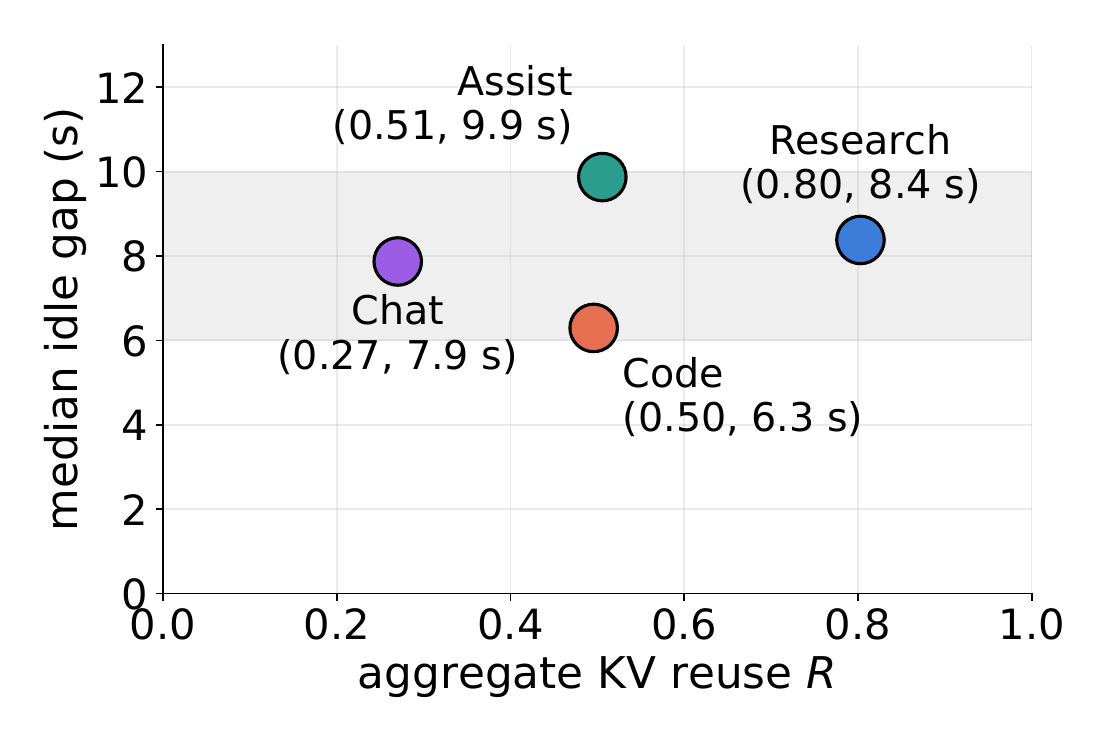}
    \caption{Aggregate KV reuse $R$ (Eq.~\ref{eq:reuse-agg}) versus the median within-session idle gap (time between consecutive model calls, intra- and inter-turn gaps pooled) for the four workloads; the shaded band spans 6--10~s.
    \Code{}'s $R$ uses its session-cumulative \texttt{cache\_read} accounting, and its idle value is its median tool-execution cadence; the other traces use provider-reported cached tokens.
    }
    \label{fig:teaser_regime}
  \end{minipage}
\end{figure}

This distinction changes the \textit{state} that an LLM serving system must manage. Within a turn, a chat application typically returns control to the user after a short orchestration path, whereas an agentic application may re-enter the model many times without human input (Figure~\ref{fig:autonomy-flow}); a session repeats this loop once per turn. Each model call first \emph{prefills} its input context, computing the attention keys and values of every context token in a \textit{Key-Value (KV) cache}~\cite{kwon2023pagedattention}, and then \emph{decodes} output tokens one at a time. Keeping reusable KV resident in GPU memory lets a later call skip prefill over a context prefix that an earlier call already processed, so the retained state remains valuable across otherwise idle intervals. Serving must therefore manage the size, lifetime, and placement of a session's evolving KV working set, in addition to scheduling the prefill and decode phases of individual model calls.

Agentic applications are emerging as a substantial component of datacenter LLM workloads, but their long-running control loops create KV residency and placement demands that request-level measurements do not capture. Recent studies (TraceLab~\cite{zhu2026tracelab}, the GitHub Copilot study~\cite{liu2026copilot}, and the ReAct study~\cite{yuan2026workload}) show that agents repeatedly invoke models and tools, carry growing contexts, and produce comparatively short outputs. None exposes the \textit{serving deployment} selected for each call, and only the Copilot study operates at \textit{production scale}, for a single coding workload and at the granularity of logical model endpoints. At hyperscaler scale, even modest errors in estimating prefix reuse, KV residency lifetime, or placement locality can translate into \textit{terabytes of excess GPU memory or extensive context reconstruction}.

We address these gaps using four production traces from \textit{Microsoft M365}. \textit{Deep Research} (\Research{}), an agent that answers a request through a long loop of model and enterprise-search tool calls; \textit{Enterprise Assistant} (\Assist{}), a single-model productivity assistant; and \textit{Coding Agent} (\Code{}). The fourth, \textit{Enterprise Chat} (\Chat{}), also invokes tools, but its internal loops are much shorter, so it serves as the lower-agency reference (Section~\ref{sec:agency}). Together, the traces contain more than $121$ million model calls.
The primary agentic trace, \Research{}, contains $17$M model calls in $691$k sessions over eight days in June 2026; \Chat{} contains $104$M model calls in $13.2$M sessions over 24 hours in July 2026 (Table~\ref{tab:traces}); a session averages $2.62$ user turns in \Research{} and $2.29$ in \Chat{}. Unlike prior traces, our telemetry combines production-scale enterprise workloads with session structure, token accounting, model identity, and, for three workloads, the serving region and deployment selected for each call.

We follow an agentic session's KV state through five questions, from how the state accumulates and is reused, to how memory capacity and deployment placement affect its residency.

\textit{\textbf{Q1.}~How agentic is each workload?} Orchestration metrics place the traces on a spectrum of agency (Section~\ref{sec:traces}).

\textit{\textbf{Q2.}~Why does KV matter?} The model--tool loop makes agentic sessions deep: \Research{} has a median \textit{session depth} of 13 model calls and a P99 of 160, compared with 3 and 52 for \Chat{}. Context keeps growing across these calls, and the three agentic workloads achieve aggregate \textit{KV reuse} of $0.50$--$0.80$, compared with $0.27$ for \Chat{}, although all four have median within-session \textit{idle gaps} of $6$--$10$~s (\Code{}: tool-execution cadence; Figure~\ref{fig:teaser_regime}; Sections~\ref{sec:reuse} and~\ref{sec:idle}).
These measurements show that the agentic workloads carry \textit{substantially more reusable KV state} across idle intervals with similar median durations
(Section~\ref{sec:session}). %

\textit{\textbf{Q3.}~Why not keep all of it?} Under the reference fp16 grouped-query-attention (GQA) KV representation (Section~\ref{sec:kvbytes}), a single session's KV can approach the memory of an entire GPU; holding every live \Research{} session at its peak working set would require an estimated $501.8$~TB; the state sits idle across heavy-tailed gaps; and its reuse is associated with staying on the serving endpoint, which we call a \emph{deployment}, that built it (Section~\ref{sec:hurdles}).

\textit{\textbf{Q4.}~What can be done?} Reclaiming state after one idle hour lowers the \Research{} hold-all-live ceiling from $501.8$ to $140.8$~TB (Section~\ref{sec:residencyworthy}), and replaying residency under constrained memory separates \textit{recomputation} caused by insufficient GPU memory from that caused by \textit{routing} when KV is deployment-local, and quantifies the potential value of \textit{affinity-aware routing}~\cite{srivatsa2024preble,pan2025kvflow,liu2025lmcache} and \textit{multi-resolution KV retention}~\cite{liu2024kivi,hooper2024kvquant,liu2024cachegen,yao2026vericache} (Section~\ref{sec:capacity}).

\textit{\textbf{Q5.}~How can others study this?} Because the traces cannot be released, we plan to release a parametric trace generator (Section~\ref{sec:agentic-trace-sim}).

\paragraph{Contributions.}
This \textit{operational systems} article makes five contributions, one per question.
\begin{enumerate}[leftmargin=*,nosep]
  \item We present a production-scale characterization of \textit{four Microsoft M365 workloads} spanning three agentic applications and one enterprise-chat workload, and compare their agency where the telemetry permits. The traces contain more than $121$ million model calls and expose session, token, and model structure and, for three workloads, deployment structure (Section~\ref{sec:traces}).

  \item We \textit{characterize} how persistent model-tool loops shape session depth, context growth, and KV reuse. Context continues growing across deep sessions, reuse is bimodal at the call level, and in \Research{} call-context KV concentrates in a predictable \textit{model class}: the heavyweight \Spine{} class, which carries the long-running context, holds $67\%$ of call-context KV bytes but only $32\%$ of uncached-context tokens and has a self-transition probability of $0.97$ (Section~\ref{sec:session}).

  \item We show why retaining every session's KV is a fleet-wide GPU-memory capacity problem. Under the reference KV representation, a single P99 \Research{} session requires $81.6$\,GiB, exceeding the 80\,GB high-bandwidth memory (HBM) of an NVIDIA H100 GPU, and holding every live \Research{} session at its peak lifetime working set requires an estimated $501.8$~TB at the exact peak. In \Research{}, mean same-model reuse is also associated with deployment continuity: $88.9\%$ when consecutive calls remain on one deployment versus $29.0\%$ after a same-model deployment change within the region (Section~\ref{sec:hurdles}).

  \item We identify two distinct sources of \textit{KV recomputation}: eviction under limited memory and unavailability under deployment-local placement.
  At a pooled memory budget of $10\%$ of the busiest-hour current-context peak (the KV that hour needs with no eviction), even clairvoyant eviction by Belady's furthest-next-use rule (FNU) leaves $31.1\%$ of \Research{} calls to recompute their context, while int4 multi-resolution retention (MRR), which keeps an evicted session's KV as a 4-bit copy instead of discarding it, reduces this call-level miss rate to $13.6\%$.
  Separately, modeling KV as \textit{deployment-local} raises the first-touch floor, the share of calls that miss even with unlimited memory, from $5.4\%$ to $20.2\%$, motivating both recoverable KV retention and affinity-aware routing (Section~\ref{sec:capacity}).

  \item Because the production telemetry cannot be released without risking client confidentiality, we plan to release \textit{\AgenticTraceSim{}}, a parametric trace generator, together with a residency replay; the generator reproduces the principal \Research{} distributions, their dependencies, and the downstream policy ordering without exposing production events (Section~\ref{sec:agentic-trace-sim}).
\end{enumerate}

\section{Agentic Workloads and How We Measure Them}
\label{sec:gaps}\label{sec:traces}

Before examining KV behavior, we establish what prior characterizations leave unmeasured, describe the four production workloads and how we measure them, and show how agentic each workload is.

\subsection{What Prior Characterizations Miss}
\label{sec:prior-misses}

Existing LLM serving systems primarily optimize request-level prefill, decode, and latency~\cite{kwon2023pagedattention,agrawal2024sarathi,zhong2024distserve}. Agentic applications instead compose repeated model and tool calls into persistent sessions whose \textit{KV cache} remains valuable between model calls~\cite{lin2024parrot}.
The closest agentic workload characterizations cover coding-agent traces and benchmark-driven agents~\cite{zhu2026tracelab,liu2026copilot,yuan2026workload}. They establish that agentic workloads contain repeated model and tool calls, growing contexts, short outputs, and substantial prefix reuse. None jointly examines diverse production enterprise workloads, the serving deployment of each call, byte-valued KV residency, and capacity-constrained residency replay at fleet scale (Table~\ref{tab:vsprior}).

These studies record the logical model of each call but not the serving \textit{deployment} selected for it. Consequently, they cannot measure how KV reuse changes when consecutive calls to the same model are served by different deployments. Our telemetry includes the serving region and deployment for three traces, allowing us to compare same-deployment continuations with same-model calls routed to another deployment (Section~\ref{sec:placement}).
Prior characterizations also report demand primarily in tokens. We convert peak session context to \textit{KV bytes} using an explicit, linearly rescalable model assumption and aggregate these estimates across concurrently live sessions (Sections~\ref{sec:kvbytes} and~\ref{sec:holdalllive}). This enables comparisons with \textit{GPU memory capacity} and exposes the sensitivity of fleet estimates to how long a session counts as live and to temporal resolution.
TraceLab evaluates cache timeouts against hit rate and relative storage, but prior studies do not compare destructive eviction policies, which discard an evicted session's KV, using byte-valued session state. We therefore develop a \textit{trace-driven residency replay} that consumes the observed production calls under fixed memory budgets (Sections~\ref{sec:coldmiss}--\ref{sec:placement-capacity}). It separates recomputation caused by limited memory from that caused by deployment-local placement, and evaluates retaining displaced KV at lower precision.

\begin{table}[t]
\caption{Features of closest agentic workload characterizations.} \label{tab:vsprior}
\footnotesize \setlength{\tabcolsep}{4pt}
\begin{tabular}{@{}L{3.5cm}cC{2.5cm}cC{3cm}@{}} \toprule
\textit{Feature} & \textbf{TraceLab}~\cite{zhu2026tracelab} & \textbf{Copilot}~\cite{liu2026copilot} & \textbf{ReAct study}~\cite{yuan2026workload} & \textbf{This work (M365)} \\ \midrule
Workload domain & Coding & Coding & Benchmarks & Enterprise research, assistant, coding, chat \\
Trace scale & $4{,}265$ sessions & $761$M calls/$13.5$M sessions & 5 benchmarks & $121$M calls/$13.9$M sessions \\
Fleet-scale production trace & -- & $\bullet$ & -- & $\bullet$ \\
Multiple agent/chat workloads & -- & -- & $\bullet$ & $\bullet$ \\
Per-call cache accounting & $\bullet$ & $\bullet$ & $\bullet$ & $\bullet$ \\ Serving deployment per call & -- & -- & -- & $\bullet$ \\
KV residency in bytes & -- & -- & -- & $\bullet$ \\ Capacity-constrained replay & -- & -- & -- & $\bullet$ \\
Released trace or generator & $\bullet$ & -- & -- & Planned \\ \bottomrule
\end{tabular}
\end{table}

\subsection{Four Production M365 Workloads}
\label{sec:fourtraces}

Table~\ref{tab:traces} summarizes the four traces, collected between late June and early August 2026. We call a workload \emph{agentic} when a user request can initiate an extended internal loop of model and tool calls without human input between calls; agency is a matter of degree, which Section~\ref{sec:agency} measures.
For each \textit{model call} $i$, the available telemetry includes \textit{context tokens} $x_i$, \textit{cached tokens} $c_i$, \textit{output tokens} $o_i$, \textit{timestamp} $t_i$ and \textit{model identity}; additional timing and tool-activity fields are available for a subset of traces.
The analyzed telemetry excludes user request/response content and contains only the operational metadata required for the measurements reported in this paper.

\paragraph{Calls, turns and sessions.}
A \emph{model call} is one inference request to an LLM.
A \emph{turn} pairs one user request with the application's complete response: its first model call is \emph{user-initiated}, and each later call is \emph{internal}, issued by the orchestrator (for example, after a tool returns) without new human input. A \emph{session} is the ordered sequence of model calls that share a session identifier; it spans one or more turns, and its \emph{depth} counts its model calls. The \emph{idle gap} before a call is the time since the preceding call of its session; it is \emph{inter-turn} if the call opens a turn and \emph{intra-turn} otherwise. \Research{} and \Chat{} log a usable turn counter; \Assist{}'s counter is empty, and \Code{} logs tool and session events rather than model calls.

For \Research{}, \Assist{}, and \Chat{}, the telemetry also identifies the serving region and endpoint instance selected for each model call. We call this endpoint instance a \emph{deployment}; multiple deployments may serve the same model. \Code{} does not expose region or deployment information.

The traces differ in granularity and cache accounting. \Research{}, \Assist{}, and \Chat{} report model calls with per-call context and provider-cached token counts. %
\Code{} contains tool executions, command-line runs, and other session events rather than model calls; its token counts are cumulative over the session, and it has an Anthropic-style \texttt{cache\_read} field. We therefore measure \Code{} depth in events, report its reuse separately, and exclude it from the byte-valued working-set and placement analyses.

\begin{table}[t]
\caption{Four production Microsoft M365 traces. Turns and internal calls are defined in Section~\ref{sec:fourtraces}; ``--'' marks traces without a usable turn counter. $R$ is aggregate reuse (Eq.~\ref{eq:reuse-agg}) over calls with a session identifier; \Code{} reports session-cumulative \texttt{cache\_read} tokens instead.
}
\label{tab:traces} \footnotesize \setlength{\tabcolsep}{3pt}
\begin{tabular}{@{}L{2.5cm}C{2.6cm}C{2.6cm}C{2.6cm}C{2.6cm}@{}} \toprule
 & \textbf{\Research{}} & \textbf{\Assist{}} & \textbf{\Code{}} & \textbf{\Chat{}} \\
 & \textit{Deep Research} & \textit{Enterprise Assistant} & \textit{Coding Agent} & \textit{Enterprise Chat} \\ \midrule
Window & 8 days & 8 days & 34-day event span & 24 hours \\
Model Calls & 17{,}008{,}715 & 217{,}106 & 1{,}119{,}432 \textit{events} & 104{,}013{,}216 \\
Sessions & 691{,}293 & 23{,}538 & 8{,}157 & 13{,}184{,}080 \\
Models & Multiple models & Single model & Multiple models & Multiple models \\ \midrule
User turns per session & 2.62 & -- & -- & 2.29 \\
Internal calls & 89.3\% & -- & -- & 59.9\% \\
Model calls per user turn & 9.38 & -- & -- & 2.49 \\
Median session depth & 13 calls & 1 call & 14 events & 3 calls \\
Realized reuse $R$ & 0.803 & 0.506 & 0.496 & 0.270 \\ \bottomrule
\end{tabular}
\end{table}

\subsection{Measurement Methodology}
\label{sec:processing}

Unless stated otherwise, descriptive measurements use the complete traces. Reuse and session-level analyses exclude calls without a valid session identifier, retaining essentially all \Research{} calls but only $75.22$M of the $104$M \Chat{} calls; the \Assist{} trace in Table~\ref{tab:traces} already contains only calls with a session identifier. The capacity-policy experiment replays the busiest one-hour \Research{} interval ($176$k calls), while the fleet-residency and idle-cap analyses (Sections~\ref{sec:holdalllive} and~\ref{sec:residencyworthy}) use the complete eligible \Research{} trace. We report the applicable population whenever it differs from the trace total.
We parse ISO-8601 timestamps directly into elapsed seconds, order calls by timestamp within each session, and remove exact duplicate \Code{} events before analysis. \Code{} samples sessions selected on June 30, 2026, and retains their complete histories, so its events span 34 days.

We distinguish three kinds of results. \emph{Measured} quantities are computed directly from production telemetry, including token counts, reuse, timestamps, and placement transitions. \emph{Estimated} quantities convert token counts to KV bytes using an explicit KV representation assumption. \emph{Replayed}
quantities report outcomes from applying counterfactual residency policies to the observed call stream.

\paragraph{KV representation assumption.}\label{sec:kvbytes}

To avoid disclosing confidential details of the deployed models, we convert tokens to bytes using a \textit{public stand-in}:
the KV geometry of the released Llama-2-70B model~\cite{touvron2023llama2}, a Grouped-Query Attention (GQA) model~\cite{ainslie2023gqa} with 80 layers, eight KV heads, and a head dimension of 128. Separate key and value tensors in fp16 give $\beta=327{,}680$ bytes/token, which we apply to every model. Absolute byte estimates (working sets, fleet ceilings, budgets, and transfer volumes) scale linearly with $\beta$, while reuse, context share, temporal ratios, and reconstruction rates at budgets set as fractions of a measured peak are independent of $\beta$.

\paragraph{Reuse.}\label{sec:metrics}The serving system keeps a \emph{prefix cache}: KV computed by earlier calls, indexed by the token prefix that produced it~\cite{vllm2026apc}; $c_i$ counts the leading context tokens of call $i$ found there. For a call with $x_i>0$, \textit{per-call reuse} is the provider-reported cached fraction of its context, \ileq{$r_i=c_i/x_i$, Eq.}{eq:reuse-percall}, and \textit{aggregate reuse} is the token-weighted fraction \ileq{$R=\sum_i c_i\big/\sum_i x_i$, Eq.}{eq:reuse-agg}. Because $c_i$ records cache behavior realized by the deployed infrastructure, reuse is a serving outcome rather than an intrinsic workload property. We define every other metric where it is first used.

\subsection{How Agentic Is Each Workload?}
\label{sec:agency}\label{sec:depth-internal}

In \Research{}, over $89\%$ of classified model calls are issued internally by the application, not the user. For \Research{} and \Chat{}, the logged \textit{turn counter} lets us measure this directly (Table~\ref{tab:traces}): a \Research{} user turn triggers far more model calls than a \Chat{} turn, and a larger share of \Research{} calls occur without a new user turn. \Research{} depth is therefore generated primarily by the application's internal orchestration rather than by additional user turns.

\Chat{} shows the same behavior to a lesser degree: it invokes tools, routes turns through multiple models, and a majority of its calls occur without a new user turn. Its loops are, however, much shorter (fewer model calls per user turn), and its sessions are shallower (Section~\ref{sec:depth}). We therefore treat \Chat{} as the lower-agency reference, not as a non-agentic, single-response chat baseline.
The event-level \Code{} trace provides complementary evidence: $84.2\%$ of its events are explicit tool executions, and the median session contains 10 tool calls. %
Neither figure is a model-to-tool call ratio. \Assist{} lacks a usable turn counter and tool field, so its internal orchestration is not measured directly; its depth distribution indicates a mixture of single-call sessions and longer multi-call sessions (Section~\ref{sec:depth}), and as a single-model workload it later serves as a placement control.

Figure~\ref{fig:teaser_regime} adds a serving outcome: the three agentic workloads realize much higher reuse than \Chat{}, while all four idle for similar median intervals between calls (\Code{}: tool-execution cadence). The next section asks what the agentic loop does to a session's KV state.

\subsection{Use of generative AI}
Generative AI tools were used to refine language and presentation, and to help implement code for trace-analysis, figure-generation, claim-verification, trace-generator and residency-replay.
They were not used to generate the study's ideas, telemetry or conclusions. The authors reviewed all assisted edits and source code, and accept responsibility for the manuscript and its artifacts.

\subsection{Privacy Considerations}
We analyze only aggregate operational metadata, excluding prompts, responses, retrieved content and other message content, and report no production identifiers. Confidential production traces are not released, and the public artifact uses only aggregate fitted parameters and synthetic traces.

\section{The Agentic Loop Makes KV Retention Valuable}
\label{sec:session}\label{sec:loop}

This section measures how persistent orchestration shapes session depth, context growth, token volume, model-class structure, and KV reuse. Within each turn, agentic sessions re-enter the model repeatedly without human input and accumulate context across these calls, which creates sustained demand for reusable KV state.

\subsection{Agentic Sessions Are Deep}
\label{sec:depth}

\begin{figure*}[t]
\centering
\subfloat[Session-depth CCDF, linear axes.\label{fig:depth}]{\quad\includegraphics[width=0.26\textwidth]{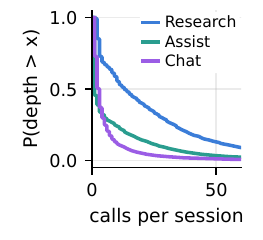}\quad} \hfill
\subfloat[Session-depth CCDF, log--log axes.\label{fig:depth_loglog}]{\quad\includegraphics[width=0.26\textwidth]{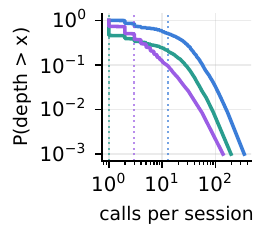}\quad} \hfill
\subfloat[\Research{} context by call index.\label{fig:context_growth}]{\quad\includegraphics[width=0.26\textwidth]{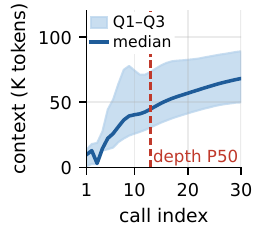}\quad}
\caption{Agentic sessions are deep and keep accumulating context.}
\label{fig:session-structure}
\end{figure*}

Figure~\ref{fig:session-structure}(a,b) shows the distribution of \textit{session depth}, the number of model calls in a session, summed over all of its turns; in \Research{}, this depth comes primarily from internal orchestration (Section~\ref{sec:agency}). \Research{} sessions have a median depth of 13 calls and a P99 of 160, compared with 3 and 52 for \Chat{}. %
\Assist{} has a median of one call but a P90 of 28, a mixture of single-call sessions and longer multi-call sessions. \Code{} counts logged events, mostly tool executions, so we report it separately and do not plot it: its median session has 14 events and its P99 session 963.

\subsection{Context Accumulates Across the Loop}
\label{sec:compaction}

\textit{Deeper sessions tend to require more KV memory} because context continues growing across model calls.
Figure~\ref{fig:session-structure}(c) shows that the median \Research{} context grows from $9{,}764$ tokens on the first call to $41{,}045$ by call 11 and $68{,}118$ by call 30, and first reaches $95\%$ of its call-30 value at call 27. Thus, estimating session demand from its first few calls would systematically understate the KV required by deep sessions.

Individual sessions, however, do not grow monotonically. In \Research{}, $90.9\%$ of multi-call sessions contain at least one \textit{context reduction}, a call with less context than its predecessor, and the median reduction leaves $23.7\%$ of the preceding call's context. The rising median in Figure~\ref{fig:session-structure}(c) therefore reflects net growth across repeated growth-and-reduction cycles.
A reduction may reflect \textit{compaction, summarization, truncation, or a transition between model roles}; the telemetry does not record its cause. Compaction near the model's context-window limit is a plausible cause, but we do not equate every reduction with an explicit compaction event.

\subsection{Context, Not Generation, Dominates Token Volume}
\label{sec:prefillbound}

We measure the \textit{context share}, the fraction of input-plus-output token volume contributed by context, as \ileq{$\rho=\sum_i x_i\big/\sum_i(x_i+o_i)$, Eq.}{eq:prefill}.
Across all four traces, \textit{context tokens} constitute at least $98\%$ of combined input and output token volume processed by model calls, with \textit{generated output tokens} accounting for less than $2\%$. The context shares are $\rho=0.98$ for \Research{}, $0.99$ for \Assist{} and \Chat{}, and $0.997$ for \Code{}. Equivalently, aggregate context-to-output ratios range from $52{:}1$ for \Research{} to $357{:}1$ for \Code{}.

Context share counts cached tokens, so it measures the KV state carried by model calls, not the prefill compute spent on uncached context; the phase comparison of the ReAct study~\cite{yuan2026workload} instead counts recomputed work, so the two are not interchangeable. \textit{Memory demand} depends on how much context remains resident, whereas \textit{recomputation} depends on how much of it has been evicted when the session resumes.

\subsection{The Loop Is Multi-model, and One Class Holds Most Reusable State}
\label{sec:class}\label{sec:modelmap}

A single \Research{} session may invoke several models. We group them into four functional classes: a heavyweight \Spine{} class that carries the long-running context, a \Mid{} class for orchestration and general reasoning, a \Scoped{} class for bounded subtasks, and a \Fanout{} class containing all remaining models.
The classes describe model roles within a session; they do not imply that every lighter-model call comes from a distinct subagent. The telemetry carries no class labels, so we assign each call to a class by matching substrings of its model name; we do not disclose the model names. The \Spine{}--\Mid{} boundary is a choice: our default \emph{conservative} mapping places only the deep-research models in \Spine{} and the reasoning models in \Mid{}, whereas a \emph{broader} mapping also places the reasoning models in \Spine{}. We report both.

\subsubsection{Call-context KV concentrates on the \Spine{}.}
\label{sec:bytes-spine}
We measure each class's share of \emph{call-context KV}: every call's context, summed over the class's calls, so a context carried by ten calls counts ten times; because one $\beta$ applies to every model, byte and token shares coincide. Figure~\ref{fig:class-structure}(a) shows that call-context KV is highly concentrated. Under the conservative mapping, the \Spine{} carries $67\%$ of call-context KV bytes, followed by \Mid{} at $31\%$; \Scoped{} and \Fanout{} together carry less than $2\%$. The broader mapping raises its share to $90\%$. The class boundary is definition-dependent, but one heavyweight class dominates \Research{} call-context KV under both mappings.

\subsubsection{Call-context bytes and \texorpdfstring{uncached prefill}{uncached prefill} have different class distributions.}
\label{sec:residency-compute}
The \Spine{} dominates call-context bytes, not uncached prefill.
Under the conservative mapping, it carries $67\%$ of call-context KV bytes but only $32\%$ of uncached-context tokens, $\sum_i \max(x_i-c_i,0)$, the context the serving system had to prefill (Figure~\ref{fig:class-structure}(b)), so most \Spine{} context is served from the prefix cache rather than recomputed.
This asymmetry makes the \Spine{} a \textit{high-value target} for residency management: it occupies substantial GPU memory, and losing its cached prefix would turn a large reused context into costly recomputation.

\subsubsection{Model-class transitions are predictable.}
\label{sec:transitions}
\begin{figure*}[t] \centering %
\subfloat[Call-context KV byte share.\label{fig:class}]{\includegraphics[width=0.33\textwidth]{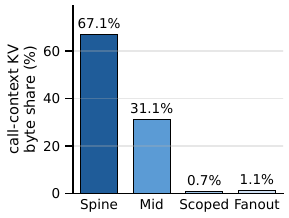}} \hfill \subfloat[\Spine{} call-context and uncached-context shares.\label{fig:residency_compute}]{\quad\includegraphics[width=0.25\textwidth]{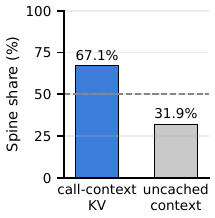}\quad} \hfill \subfloat[Model-class transition probabilities.\label{fig:transition}]{\includegraphics[width=0.35\textwidth]{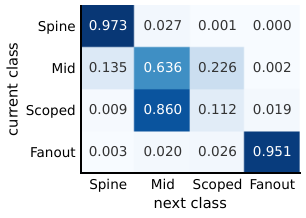}}  \caption{Call-context KV concentrates on a predictable model class. (a) Under the conservative mapping, \Spine{} carries $67.1\%$ of call-context KV bytes; the broader mapping raises this share to $90.0\%$. (b) \Spine{} accounts for only $31.9\%$ of uncached-context tokens. (c) \Spine{} is strongly self-transitioning, with $P_{\Spine\Spine}=0.97$; selecting the most frequent outgoing transition predicts the next class for approximately $88\%$ of observed transitions.} \label{fig:class-structure} \end{figure*}

Consecutive calls within a session also follow a strong class-transition structure (Figure~\ref{fig:class-structure}(c)). The \Spine{} and \Fanout{} self-transition with probabilities $0.97$ and $0.95$, \Scoped{} transitions to \Mid{} with probability $0.86$, and \Mid{} either remains in its class or transitions to \Scoped{} or \Spine{}. Predicting the next class as the most frequent successor of the current class is correct for $\approx 88\%$ of observed \Research{} transitions.
This low-cost signal indicates which model, and hence which KV state, a session is likely to access next. The opportunity is \textit{workload-dependent}: it exists in multi-class \Research{} but not in single-model \Assist{}.

\subsubsection{Reuse declines after model-class excursions.}
\label{sec:excursion}
\begin{figure}[t]
  \centering
  \begin{minipage}[t]{0.49\linewidth}
    \centering
    \quad\includegraphics[width=0.73\linewidth]{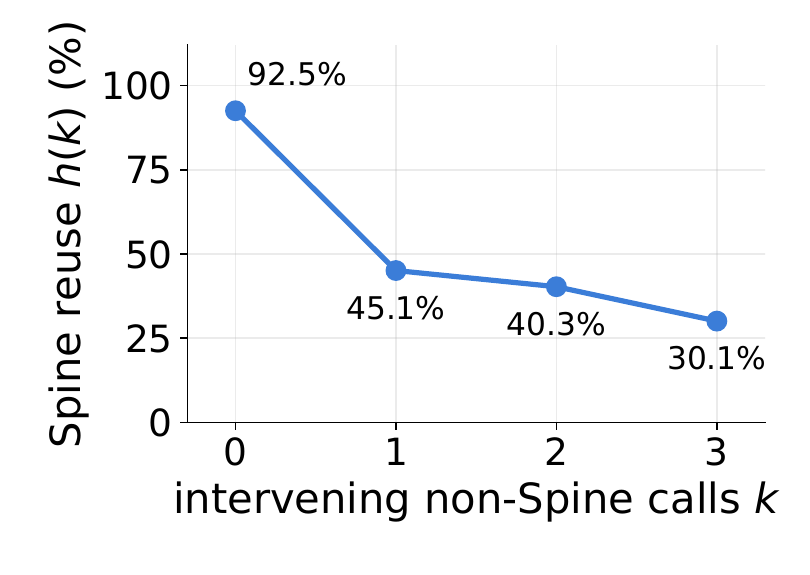}\quad
    \caption{\Spine{} reuse declines after an excursion. Mean reuse falls from $92.5\%$ for consecutive \Spine{} calls to $45.1\%$ after one intervening non-\Spine{} call. The analysis does not independently control for elapsed time or deployment placement.}
    \label{fig:excursion}\label{fig:lifecycle}
  \end{minipage}\hfill
  \begin{minipage}[t]{0.49\linewidth}
    \centering
    \quad\includegraphics[width=0.8\linewidth]{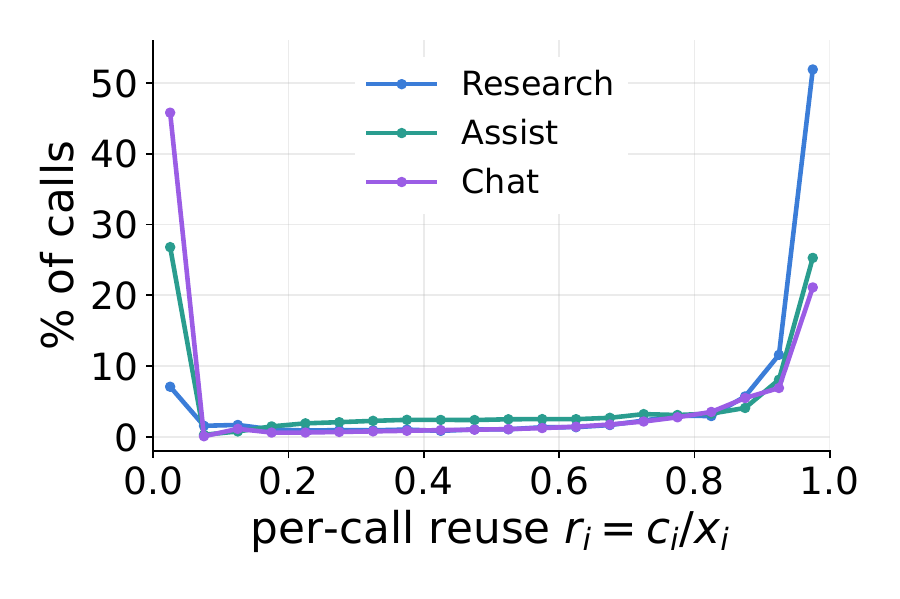}\quad
    \caption{Per-call KV reuse is bimodal, but differently weighted. \Research{} is warm-weighted, with $51.9\%$ of calls achieving at least $95\%$ reuse, whereas \Chat{} is cold-weighted, with $47.6\%$ of calls below $20\%$ reuse. \Assist{} has the largest intermediate-reuse share.}
    \label{fig:reuse}
  \end{minipage}
\end{figure}

Model-class control flow is also associated with whether retained \Spine{} state remains reusable. For each \Spine{} call $i$ after the first in its session, let $k$ be the number of non-\Spine{} calls since the preceding \Spine{} call (the \textit{excursion} length), and define \ileq{$h(k)=\mathbb{E}[r_i\mid i\in\Spine,\ k]$, Eq.}{eq:excursion}.
Mean reuse is $92.5\%$ for consecutive \Spine{} calls ($k=0$), falls to $45.1\%$ after one intervening call, and is $30.1\%$ after three or more (Figure~\ref{fig:excursion}). Most of the decline comes with the first intervening call.
Model-class history is therefore a candidate signal, alongside wall-clock age (the idle time examined in Section~\ref{sec:idle}), for estimating the future value of resident KV state. Because the analysis does not control for elapsed time or deployment placement, the decline is an association, not evidence of a causal effect of model-class interference; Section~\ref{sec:placement} examines how deployment continuity relates to reuse.

\subsection{Reuse Is Bimodal at the Call Level}
\label{sec:reuse}

\textit{KV reuse} lets a model call skip recomputing context that earlier calls in its session already processed, whether the call continues a turn or opens a new one. Figure~\ref{fig:reuse} shows that per-call reuse $r_i$ is bimodal: calls concentrate near low or high reuse, but the weights of the cold and warm modes differ across workloads.
\Research{} is warm-weighted, with $51.9\%$ of calls having at least $95\%$ reuse and only $11.4\%$ having reuse below $20\%$. \Chat{} is cold-weighted: $47.6\%$ of calls have reuse below $20\%$, while only $21.1\%$ reach $\geq 95\%$. \Assist{} is more evenly distributed, with $29.9\%$ of calls in the intermediate $20$--$80\%$ range.
The bimodality is a \textit{per-call property}: only $40.7\%$ of \Research{} sessions have a session-level aggregate reuse of at least $0.8$, and $52.1\%$ lie between $0.2$ and $0.8$, because sessions mix cold calls with warm continuations. Prefix-level reuse is therefore a useful residency signal, but a policy must combine it with session continuation, working-set size, and fleet memory pressure.

\textit{Aggregate reuse} $R$ is $0.27$ for \Chat{}, $0.50$ for \Code{}, $0.51$ for \Assist{}, and $0.80$ for \Research{}. The high \Research{} value reflects repeated access to accumulated context, but realized reuse also depends on the \textit{serving system's} ability to retain and recover that context; Section~\ref{sec:placement} shows that reuse is strongly associated with whether consecutive calls remain on the same deployment.
Equivalently, \Research{} has a \textit{context-carry tax} $\tau=1/(1-R)$ of $5.1$: %
it processes $5.1$ context tokens for every uncached context token it must prefill, and the rest is served from the prefix cache. Losing that cached context would multiply aggregate context prefill work by up to $5.1\times$, which is why KV retention matters. The tax is close to TraceLab's prefill amplification~\cite{zhu2026tracelab} ($5.1\times$ versus $5.3\times$), although the denominators differ.

Retained KV is therefore valuable: agentic sessions are deep, their context keeps growing, much of it is served from the prefix cache, and in \Research{} most call-context KV bytes sit in one large, predictable \Spine{} state. Why, then, not keep every session's KV resident for the session's lifetime? Section~\ref{sec:hurdles} examines the hurdles this policy faces.

\section{Why Not Keep Every Session's KV Resident?}
\label{sec:hurdles}

The simplest residency policy retains each session's KV state from its first to its last call. We now show four hurdles to this policy: a single session can fill a GPU's memory, live sessions add up across the fleet, the retained state sits idle across heavy-tailed gaps, and its reuse is associated with staying on the deployment that built it.

\subsection{A Single Session Can Fill a \texorpdfstring{GPU}{GPU}}
\label{sec:persessionws}\label{sec:workingset}

Under the reference KV representation (Section~\ref{sec:kvbytes}), the estimated \textit{per-session KV working set} is the peak KV state of session $s$, \ileq{$W_s=\beta\cdot\max_{i\in s}(x_i)$, Eq.}{eq:workingset}. This is a byte estimate, not a measurement of observed physical HBM allocation.

\begin{figure}[t]
\begin{minipage}{0.49\linewidth}
  \centering
\quad\includegraphics[width=0.87\linewidth]{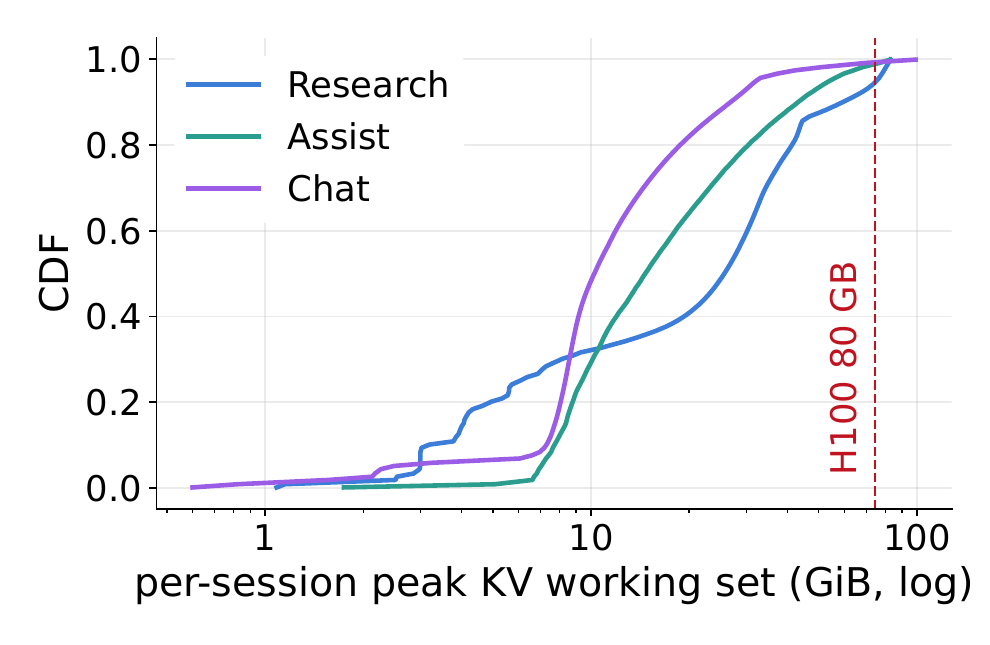} \quad
 \caption{Per-session KV working sets differ across workloads. The dashed line marks the 80~GB ($74.5$~GiB) HBM of an H100.
}
\label{fig:working_set}
  \end{minipage}\hfill
  \begin{minipage}{0.49\linewidth}
    \centering
    \quad\includegraphics[width=0.87\linewidth]{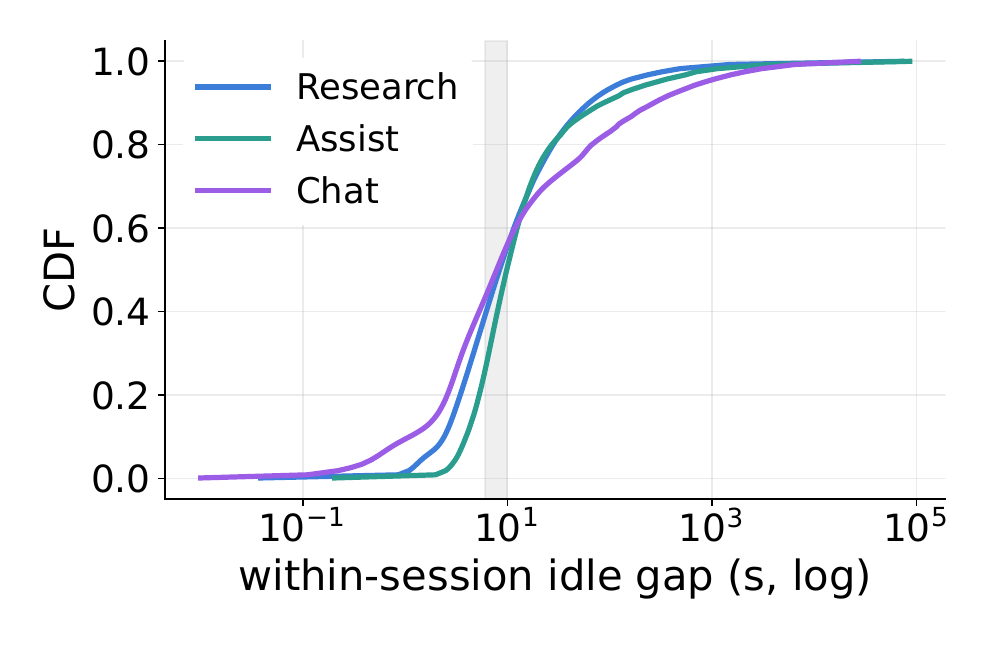}\quad
    \caption{Gaps between consecutive model calls of a session (intra- and inter-turn pooled) are heavy-tailed: medians are $7.9$--$9.9$~s, but the tail extends to minutes or days.}
    \label{fig:idle}
  \end{minipage}
\end{figure}

A \Research{} session requires an estimated $25.7$~GiB at P50, $59.2$~GiB at P90, and $81.6$~GiB at P99 (Figure~\ref{fig:working_set}). These estimates include KV state alone: model weights, activations, and runtime state require additional GPU memory.
For scale, the P90 working set is comparable to most of the 80~GB HBM of one H100, while the P99 estimate exceeds this capacity. In practice, a model with the reference attention configuration may be served by a tensor-parallel group of GPUs, each holding the KV of a subset of the attention heads~\cite{kwon2023pagedattention};
the tensor-parallel model replica, rather than one GPU, must collectively hold the session's KV.
The working-set distributions also differ across workloads (Figure~\ref{fig:working_set}): a median \Research{} session carries approximately $2.5\times$ the KV state of a median \Chat{} session.

\subsection{Across the Fleet, Live Sessions Add Up}
\label{sec:holdalllive}
Across the serving fleet, the large working sets of concurrent, long-running agentic sessions place sustained pressure on scarce and costly HBM.
We bound this pressure with a deliberately pessimistic \textit{hold-all-live} model in which each session retains its peak lifetime KV working set from its first through its last observed call. Let $\mathcal{S}(t)=\{s \mid t_s^{\mathrm{first}}\leq t\leq t_s^{\mathrm{last}}\}$ denote the sessions live at time $t$. The estimated fleet-wide KV residency is \ileq{$K_{\mathrm{live}}(t)=\sum_{s\in\mathcal{S}(t)}W_s$, Eq.}{eq:activekv}, and the full-trace residency ceiling is $K_{\mathrm{live}}^\star=\max_t (K_{\mathrm{live}}(t))$.

Treating all regions as one logical pool, an exact event sweep of Equation~\ref{eq:activekv} over the eight-day \Research{} trace gives $K_{\mathrm{live}}^\star=501.8$~TB, and the number of concurrently live sessions peaks at $11{,}358$. This is the estimated footprint of KV state alone under the reference representation; it excludes model weights and other runtime memory.
\textit{Temporal resolution} matters for this ceiling. One-minute bins give $504.9$~TB, $0.6\%$ above the exact value, whereas hourly bins, which count a session as resident for every hour its lifetime touches, give $709.4$~TB, a $1.41\times$ overstatement. We use the exact ceiling for capacity results and report the hourly value only as a resolution warning.

\subsection{Retained State Sits Idle Across Heavy-tailed Gaps}
\label{sec:lifecycle}\label{sec:idle}

A session's KV state remains potentially reusable between model calls, but retaining it consumes scarce GPU memory while the session is idle.
Within-session idle gaps, measured between consecutive model calls of a session and pooling intra- and inter-turn gaps (Section~\ref{sec:fourtraces}), have similar medians across the three traces that report per-call tokens: $8.38$~s for \Research{}, $9.87$~s for \Assist{}, and $7.87$~s for \Chat{}. Their tails are substantially longer, with P99 gaps of approximately $20$, $50$, and $93$ minutes, respectively (Figure~\ref{fig:idle}). The longest observed \Research{} gap is $7.9$ days, indicating that some sessions are parked and later resumed.
In the two traces with a usable turn counter,
intra-turn gaps dominate the short-gap body of the distribution, while inter-turn gaps account for most of the total idle time:
in \Research{}, inter-turn gaps constitute $7.0\%$ of gaps but $92.6\%$ of total idle time (median $125.0$~s versus $7.5$~s for intra-turn gaps), and in \Chat{}, whose maximum gap is bounded by its 24-hour window, they constitute $27.4\%$ of gaps and $97.3\%$ of idle time ($146.5$~s versus $4.6$~s).

This range makes any single retention timeout a compromise: a short timeout discards KV that returning sessions would have reused, forcing reconstruction, while a long timeout holds state for sessions that stay inactive for hours or days. Section~\ref{sec:residencyworthy} quantifies this trade-off by varying the idle interval after which a session's KV state is reclaimed.

\subsection{Reuse Is Associated with Staying on One Deployment}
\label{sec:placement-sec}

The preceding analysis treats resident KV as logically available whenever a session returns. In a distributed fleet, however, reusable KV may remain local to the serving deployment that constructed it. Using the call-level region and deployment fields available for \Research{}, \Assist{}, and \Chat{}, we examine how deployment continuity is associated with reuse.

\subsubsection{Long \Research{} sessions span multiple deployments.}
\label{sec:locality}
We summarize placement continuity using single-region (SR) and single-deployment (SD) session subsets. Both are defined per model: every model a session invokes stays within one region (SR) or on one deployment (SD), although different models may use different regions or deployments.

Placement stability differs substantially across workloads. Only $45.0\%$ of \Research{} sessions are SD and $57.6\%$ are SR, compared with $80.2\%$ and $83.3\%$ for single-model \Assist{}, and $79.1\%$ and $80.0\%$ for \Chat{}. A \Research{} session touches a mean of $4.48$ deployments across $3.04$ regions, reflecting its longer, multi-model orchestration path.

The SR and SD subsets are structurally different from the complete workloads. For example, the median \Research{} depth falls from 13 calls overall to 5 in SR and 3 in SD. Their aggregate reuse and capacity values therefore cannot be interpreted as the causal effect of applying a placement policy. We use these subsets only to describe placement stability; the per-transition analysis below provides the cleaner comparison of reuse immediately following a deployment change.

\subsubsection{Reuse is strongly associated with deployment continuity.}
\label{sec:placement}
\begin{figure}[t]
  \centering
  \begin{minipage}[t]{0.5\linewidth}
    \centering
    \includegraphics[width=0.95\linewidth]{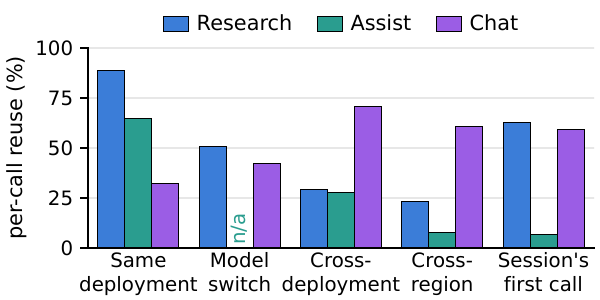}
    \caption{\textbf{KV reuse is associated with deployment continuity in the agentic traces.} For \Research{}, mean reuse falls from $88.9\%$ on same-deployment continuations to $29.0\%$ when the same model moves to another deployment. The single-model \Assist{} control shows a similar decline from $64.8\%$ to $27.7\%$. \Chat{} is selection-confounded: its same-model deployment moves are rare, and the categories can differ in session composition.}
    \label{fig:placement_reuse}
  \end{minipage}\hfill
  \begin{minipage}[t]{0.48\linewidth}
    \centering
    \includegraphics[width=0.9\linewidth]{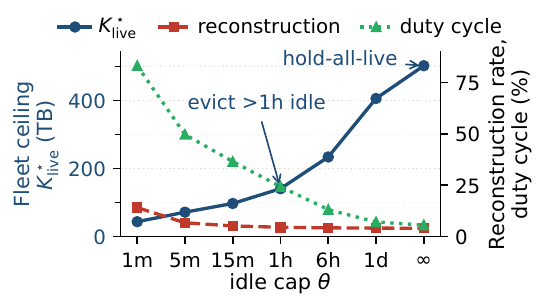}
    \caption{
    \textbf{An idle cap substantially reduces fleet KV demand.} For \Research{}, reclaiming state after one idle hour reduces the exact fleet ceiling $K^\star_{\mathrm{live}}$ from $501.8$ to $140.8$~TB while increasing reconstruction from $4.07\%$ to $4.53\%$. The dotted line (right axis) is the duty cycle, engine time as a share of resident session time. The penalty rises more rapidly when the cap enters the short-gap body of the distribution.
    }
    \label{fig:idlecap}
  \end{minipage}
\end{figure}

For each model call, we condition per-call reuse on its transition from the preceding call in the session. We distinguish a continuation on the preceding call's deployment, a model switch, a same-model move to another deployment within the region, a same-model cross-region move, and a session's first call. Figure~\ref{fig:placement_reuse} reports the resulting mean reuse.

For \Research{}, same-deployment continuations achieve mean reuse of $88.9\%$. When the same model is served by another deployment within the region, mean reuse falls to $29.0\%$; for a cross-region move, it falls to $23.5\%$. These transitions account for only $2.0\%$ and $0.9\%$ of calls, but their reuse is less than one-third that of same-deployment continuations; same-deployment reuse is stable across the eight observed days.

The single-model \Assist{} trace removes the model-switch confound. Its mean reuse is $64.8\%$ for same-deployment continuations, $27.7\%$ after a cross-deployment move, and $7.5\%$ after a cross-region move. The consistent decline in both \Research{} and \Assist{} provides evidence that deployment continuity is strongly associated with preserving reusable KV state.

\Chat{} does not show the same pattern. Model switches account for $64.0\%$ of its calls, while same-model deployment and region moves together account for less than $2.5\%$. These rare moves show higher mean reuse than same-deployment continuations, but the categories can differ in session depth and composition, so we treat \Chat{} as selection-confounded for this comparison and base the deployment-continuity finding on \Research{} and the single-model \Assist{} control.

Together, these measurements show why holding all KV state is costly: under the reference representation a tail session can exceed one GPU's memory, live sessions add up to a fleet-wide capacity problem, much of the state idles for long intervals, and its reuse is associated with the deployment continuity that routing provides. The next section evaluates what residency policies can recover under these constraints.

\section{What Can Be Done? Residency Policies Under Constrained Memory}
\label{sec:capacity}\label{sec:policies}

Because holding every session's KV state is costly, we replay residency policies on the observed call streams. %
We first sweep an idle cap over the complete \Research{} trace, then compare destructive eviction with multi-resolution retention (MRR), which keeps displaced KV as a recoverable int4 copy, under fixed budgets with one KV object per session, and finally measure the miss rate that remains with unlimited memory when KV objects are deployment-local.

A \textit{trace-driven residency replay} processes an ordered call stream $\mathcal{C}$ under a pooled memory budget $B$, shared by all sessions in the stream, and KV residency policy $\pi$: it handles calls in timestamp order and keeps the KV states that $\pi$ retains within $B$. A call whose KV state is absent when it arrives is a \emph{miss} and requires \emph{reconstruction}: its whole context must be recomputed by prefill. With $m_{\pi,B}$ misses, the replayed reconstruction (miss) rate is
\ileq{
$M_{\pi}(B)=\frac{m_{\pi,B}}{|\mathcal{C}|}$, Eq.
}{eq:coldmiss}; $M_{\mathrm{LRU}}$, $M_{\mathrm{FNU}}$, and $M_{\mathrm{MRR}}$ denote it for the policies below. The replay treats a session's current KV state as one variable-sized object, updates its size at each call, and applies $\pi$ whenever resident KV exceeds $B$. Because each replay starts with an empty cache, its first access to each KV object in $\mathcal{C}$ misses, independent of the policy and budget. This \emph{first-touch floor}, the number of distinct KV objects divided by $|\mathcal{C}|$, is the compulsory-miss rate of classical cache analysis, and every policy reaches it once $B$ is at least the stream's no-eviction peak.

\subsection{Reclaiming Idle State: Nominally Live State Is Not Always Residency-worthy}
\label{sec:residencyworthy}

The \textit{hold-all-live} ceiling is deliberately pessimistic: a session whose consecutive calls are days apart holds its peak KV state throughout the gap. We therefore ask which live sessions are \emph{residency-worthy}, likely to resume soon enough that keeping their KV resident pays off, and parameterize liveness by an \textit{idle cap} $\theta$: once a session has been idle for $\theta$, its KV state is evicted from HBM and its next model call requires reconstruction. Each session is still sized at its peak lifetime working set, and $\theta=\infty$ recovers the hold-all-live model.

Figure~\ref{fig:idlecap} shows that most of the hold-all-live ceiling can be reclaimed at little reconstruction cost. A one-hour cap lowers the \Research{} ceiling $3.6\times$, from $501.8$ to $140.8$~TB, while the replayed reconstruction rate over the complete trace rises only from $4.07\%$ at $\theta=\infty$, where only first touches count, to $4.53\%$. The long idle tail thus inflates the ceiling, yet few calls follow such gaps.
The trade-off remains favorable at a 15-minute cap, where the ceiling falls to $97.2$~TB and reconstruction reaches $5.2\%$. As the cap falls further below the approximately 20-minute P99 idle gap (Section~\ref{sec:idle}), reconstruction rises quickly, to $6.7\%$ at five minutes and $14.1\%$ at one minute. The source of this penalty shifts with the cap: inter-turn gaps (Section~\ref{sec:idle}) are $98.0$--$99.8\%$ of the gaps reclaimed at caps of 15 minutes or longer, whereas at one minute intra-turn gaps contribute $5.23$ of the $10.04$ added percentage points. \textit{Fleet provisioning} should therefore distinguish nominally live sessions from those likely to resume soon, rather than treating each session's first-to-last-call span as continuously resident.

\subsection{Capacity Limits Destructive Eviction}
\label{sec:coldmiss}\label{sec:replay}

The idle-cap analysis charges each resident session its peak lifetime working set; we next replay current-context KV residency under a fixed memory budget $B$.

\begin{figure*}[t]
\centering
\subfloat[\Research{}, busiest hour: $K^\star_{\mathrm{hr}}=47.0$~TB.\label{fig:coldmiss-researcher}]{\includegraphics[width=0.31\textwidth]{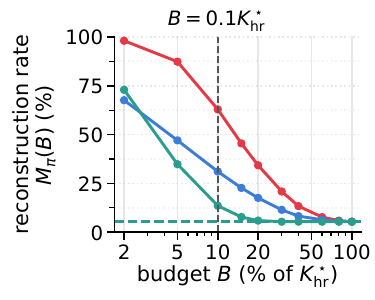}} \hfill
\subfloat[\Assist{}, full window: $K^\star_{\mathrm{win}}=8.1$~TB.\label{fig:coldmiss-assist}]{\includegraphics[width=0.31\textwidth]{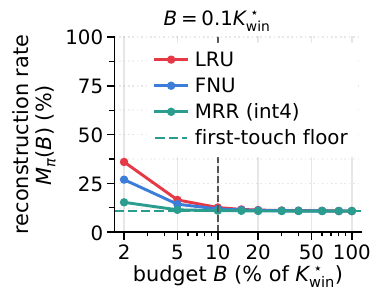}} \hfill
\subfloat[\Chat{}, busiest hour: $K^\star_{\mathrm{hr}}=1{,}876.7$~TB.\label{fig:coldmiss-chat}]{\includegraphics[width=0.31\textwidth]{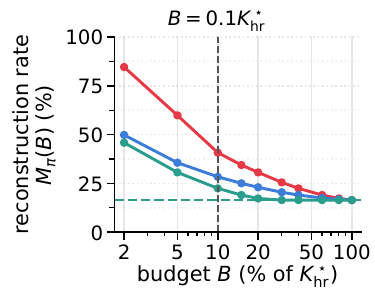}}
\caption{\textbf{Capacity constrains destructive eviction, while multi-resolution retention reduces reconstruction.} At $B=0.1K_{\mathrm{hr}}^\star$, \Research{} call-level reconstruction rates $M_\pi(B)$ are $62.9\%$ for LRU, $31.1\%$ for FNU, and $13.6\%$ for int4 MRR. \Chat{} also exhibits substantial policy separation, whereas \Assist{} does not. \Research{} and \Chat{} use their busiest-hour replay streams; \Assist{} uses its complete retained stream because of its lower volume. Budgets are pooled and expressed as fractions of each stream's no-eviction current-context peak: $K^\star_{\mathrm{hr}}$ in (a) and (c), $K^\star_{\mathrm{win}}$ in (b). MRR keeps int4 copies within the same pooled memory budget. The dashed line marks the first-touch floor (Section~\ref{sec:replay}). MRR reaches this floor when displaced KV instead moves to a host-memory tier four times the pooled memory budget (Section~\ref{sec:mrr}). The legend in (b) applies to all panels.}
\label{fig:coldmiss}
\end{figure*}

The replay processes the $176{,}007$ model calls that $9{,}486$ sessions issue in the busiest \Research{} hour, so its first-touch floor is $5.4\%$, and sizes each session by its \textit{current context} at each call rather than its lifetime peak. Without eviction, the current-context peak in this hour is $K_{\mathrm{hr}}^\star=47.0$~TB, a different quantity from the whole-trace, lifetime-peak ceiling $K_{\mathrm{live}}^\star$.
We set $B=0.1\,K_{\mathrm{hr}}^\star=4.7$~TB, a deliberately constrained operating point at which policy differences are visible, and sweep the full budget range in Figure~\ref{fig:coldmiss}. Here $B$ is a pooled budget: the replay treats the KV memory of all replicas serving the stream as one tier and selects victims across the whole stream. Physical replicas, each a tensor-parallel group of GPUs, enforce residency on their own memory, so at the same total capacity we expect per-replica caches to reconstruct more often than the pooled replay reports. Because of its lower volume, \Assist{} is replayed over its complete retained stream, whose no-eviction current-context peak we denote $K^\star_{\mathrm{win}}$ ($8.1$~TB).

We compare two \emph{destructive} policies, which discard an evicted session's KV. \textit{LRU} evicts the least recently accessed session~\cite{vllm2026apc,wang2025kvcache}.
The clairvoyant \textit{furthest-next-use (FNU)} policy applies Belady's rule~\cite{belady1966}: it knows the next access to each resident KV object and evicts the object whose next access lies furthest in the future. Belady's rule minimizes misses when all objects have the same size; because session KV objects vary in size and grow with context, FNU is here a strong clairvoyant heuristic rather than an optimum~\cite{berger2018practical}.
At $B=4.7$~TB, LRU reconstructs context for $62.9\%$ of \Research{} model calls and FNU for $31.1\%$.
Future knowledge thus approximately halves reconstruction, yet nearly a third of calls still rebuild their context.

\subsection{Retaining a Lower-precision Copy: Multi-resolution Retention}
\label{sec:mrr}

In the replay, \textit{MRR} replaces eviction with \emph{demotion}: when resident KV exceeds $B$, it retains the least recently used session as an int4 copy, a quarter of its fp16 size, which stays in HBM and is charged to the same budget.
A later call to a demoted session is modeled as \emph{promoting} the copy back to fp16 instead of reconstructing its context, building on KV quantization and compression~\cite{liu2024kivi,liu2024cachegen}; VeriCache likewise keeps a compressed KV copy close to the GPU~\cite{yao2026vericache}. MRR chooses which session to demote by recency alone, without future knowledge; only when every other session is already int4 and the budget is still exceeded does it discard int4 copies, least recently used first, and a session whose copy was discarded is reconstructed on its next call.
The replay does not measure how int4 KV changes responses, but published evaluations of open Llama models report modest costs.
Adding 4-bit KV quantization to an already weight- and activation-quantized model raises WikiText-2 perplexity by $0.12$ for Atom (W4A4 Llama-7B)~\cite{zhao2024atom} and $0.14$ for QoQ (W4A8 Llama-2-7B)~\cite{lin2025qserve}.
Group-quantized W4A8KV4 Llama-2 models from 7B to 70B lose at most $1.03$ points of FP16 average zero-shot accuracy; the 70B model has the reference KV geometry used in our byte estimates.
W4A8KV4 Llama-3.1-8B loses $0.14$ points of its BF16 LongBench average ~\cite{lin2025qserve}. These evaluations use single-request benchmarks rather than agentic sessions.

With the same pooled HBM budget, MRR \textit{reduces the replayed reconstruction rate} to $13.6\%$, $2.3\times$ below FNU. Because MRR demotes in LRU's victim order, the drop from LRU's $62.9\%$ isolates the effect of keeping a smaller, lower-precision copy instead of destroying the state, at a potentially modest quality cost. This does not contradict Belady's result, which applies to policies that keep or discard whole objects: a quarter-size copy lets the same budget hold up to four times as much session state, and MRR's rate at $B$ nearly equals LRU's at $4B$ ($13.56\%$ versus $13.35\%$). The advantage is therefore one of capacity, not foresight; at the tightest budget evaluated, $0.02\,K^\star_{\mathrm{hr}}$, FNU is lower ($67.60\%$ versus $72.92\%$; Figure~\ref{fig:coldmiss-researcher}).
Demoting displaced KV to a separate host-memory (CPU DRAM) tier, rather than keeping the int4 copy in HBM, lowers the rate further: a host tier equal to the pooled HBM budget gives $9.46\%$, and one twice as large gives $5.57\%$, close to the $5.4\%$ first-touch floor (dashed line in Figure~\ref{fig:coldmiss}) that a tier four times the budget reaches. Doing so moves $1{,}026.1$~TB of int4 KV between host and GPU memory in the hour, $285.0$~GB/s on average. The replay charges no promotion latency, dequantization, or transfer cost and treats a promoted int4 copy as equivalent to an fp16 hit. Counting per-group quantization metadata enlarges each int4 copy beyond a quarter of fp16 and raises the in-HBM MRR rate to between $14.99\%$ and $19.25\%$. These results therefore motivate retaining recoverable KV representations across memory tiers, subject to the quality, latency, and bandwidth costs the replay omits.

\subsection{Policy Separation Differs Across Workloads}
\label{sec:synthesis}

The capacity result is not uniform across workloads.
\Assist{} has comparable reconstruction rates of $12.6\%$, $11.9\%$,
and $11.1\%$ for LRU, FNU, and MRR, respectively,
all close to its $10.8\%$ first-touch floor, which leaves better policies little to recover at the 10\% pooled budget. \Chat{}, despite smaller per-session working sets, separates the three policies at $40.7\%$, $28.4\%$, and $22.5\%$, above its $16.4\%$ first-touch floor; its $9{,}749$~TB hold-all-live ceiling over 24~h comes from concurrency, not session size.
These workloads expose \textit{two routes to fleet GPU-memory pressure}: deep agentic sessions create large per-session working sets, while \Chat{} creates a large aggregate footprint through concurrency. Capacity planning must therefore account for both the \textit{distribution of session working sets} and the \textit{number of concurrently live sessions}; neither session size nor workload category alone determines whether KV residency becomes HBM capacity-bound.

\subsection{Routing: Deployment-local KV Creates an Independent Floor}
\label{sec:placement-capacity}

The replay so far models one KV object per session. To represent deployment-local state, we repeat the busiest-hour \Research{} replay with KV objects keyed by both session and deployment. Over the same $176{,}007$ calls, this increases the number of modeled cache objects from $9{,}486$ to $35{,}588$, because a session routed among deployments may establish separate state at each one.

With unlimited residency capacity, reconstruction occurs only on the first access to each modeled cache object. Session-only keying gives the $5.4\%$ first-touch floor of Section~\ref{sec:replay}, while session-and-deployment keying raises it to $20.2\%$. Deployment-local state thus imposes a routing floor that neither more memory nor better eviction within a deployment can remove; lowering it requires deployment affinity, KV replication, or migration.
This replay assumes that KV remains local to the deployment identifier reported by the telemetry and is not automatically shared across deployments.

Together, the capacity and placement results identify two distinct sources of reconstruction. Limited GPU memory displaces reusable KV even under clairvoyant FNU eviction, while deployment changes make state unavailable under deployment-local caching. Recoverable retention, such as MRR, mitigates the first; affinity-aware routing, replication, or migration mitigates the second.

\subsection{Lessons learned}
\label{sec:lessons}
Relative to prior serving work, our production measurements offer four lessons. First, request-level KV management~\cite{kwon2023pagedattention,vllm2026apc} does not consider reuse across calls within a session. In \textsc{Research}, 89.3\% of calls are internal and 80.3\% of context tokens hit the prefix cache, motivating session-level residency decisions. Second, retaining KV throughout a session's lifetime can overestimate demand: a one-hour idle cap reduces the \textsc{Research} ceiling from 501.8 to 140.8\,TB, while reconstruction increases from 4.07\% to 4.53\%. Third, eviction policy has limited impact under tight memory: clairvoyant FNU reconstructs 31.1\% of calls, while retaining an int4 copy within the same budget reduces this to 13.6\%. Finally, locality-aware routing~\cite{liu2025lmcache,pan2025kvflow,srivatsa2024preble} is relevant in production, since mean reuse falls from 88.9\% to 29.0\% after a same-model deployment change.

\subsection{Open Problems and Limitations}
\label{sec:openproblems}

Our measurements expose two complementary opportunities for agentic LLM serving: preserving reusable KV under limited GPU memory and coordinating its placement with application control flow. We frame these as research directions rather than evaluated system designs.

\paragraph{Reversible, multi-resolution KV residency.}\label{sec:op-reversible}
MRR isolates one effect, keeping displaced KV recoverable instead of destroying it; it does not show that int4 retention is the best mechanism. Residency mechanisms leave the context unchanged and decide only whether, where, and at what precision its KV is kept: discard and recompute, exact offload to host or remote storage with a fetch on resumption~\cite{qin2025mooncake,liu2025lmcache,abhyankar2024infercept}, or quantized retention as in MRR~\cite{liu2024kivi,hooper2024kvquant,liu2024cachegen}. Context reduction, such as compaction, rolling summaries, or agent memories that keep a pointer and retrieve content on demand~\cite{kang2026acon,packer2023memgpt,chhikara2025mem0}, changes the context itself: it can change the response, and because prefix caching reuses only an unchanged prefix, each edit invalidates the cached KV after it. Context reduction and KV residency can be combined: reduction shrinks $W_s$, residency retains the remaining state, and our replay already sizes sessions after the application's own reductions (Section~\ref{sec:compaction}). Exact offload avoids MRR's precision loss but moves four times the bytes of an int4 copy; it beats reconstruction only if fetch bandwidth exceeds $\beta R_{\mathrm{prefill}}$, at least about $1$~GB/s for fp16, and its latency adds to the resumed call unless the fetch is issued during the preceding idle gap. Hybrids such as VeriCache decode from a compressed copy and verify against a full-precision copy fetched from a lower tier~\cite{yao2026vericache}. The open problem is to decide, for each session, whether its state should be retained exactly, quantized, offloaded, summarized, or discarded and reconstructed, given the workload's idle gaps, reuse and quality tolerance, while meeting latency and output-quality objectives.

\paragraph{Application-aware placement and lifecycle control.}\label{sec:op-observability}
Model-class predictability, excursion-conditioned reuse (Section~\ref{sec:class}), and deployment continuity (Sections~\ref{sec:placement-sec} and~\ref{sec:placement-capacity}) motivate coordinating application control flow with KV placement and retention; a context reduction (Section~\ref{sec:compaction}) likewise lowers the session's current KV requirement and provides a natural point for releasing or demoting superseded state.
Current serving interfaces reveal little about an agent's expected next model, tool interval, context reduction, or likelihood of resumption. Prior systems expose parts of this structure through semantic variables, program-level scheduling, cache-aware routing, and application hints~\cite{lin2024parrot,luo2025autellix,gim2025pie,pan2025kvflow,agarwal2026softwaredefined}. A broader interface could allow the orchestrator to communicate model role, expected continuation, and session-lifecycle events explicitly, while the serving layer returns memory pressure and placement constraints. Designing this interface without compromising isolation, load balance, or failure tolerance remains an open problem.
Such mechanisms also require portable measurement. In our traces, three workloads report provider-cached tokens per call, while \Code{} uses session-cumulative \texttt{cache\_read} accounting. A provider-neutral definition of cache objects, KV reuse, deployment continuity, and reconstruction would enable residency policies to be compared consistently across serving stacks~\cite{openai_promptcaching,vllm2026apc,wang2025kvcache}.

\paragraph{Scope and limitations.}\label{sec:caveats}
Our results describe four Microsoft M365 workloads observed over specific windows between late June and early August 2026 and should not be interpreted as universal properties of agentic or chat serving. \textit{Cached tokens} reported by the LLM endpoint measure the reuse realized by the deployed infrastructure, whereas our byte-valued working sets scale linearly with the assumed per-token KV size $\beta$. The deployment identifier, the finest reported placement label, may not expose worker-level placement or cache sharing.
Finally, several results depend on analysis choices: the model-class boundary (Section~\ref{sec:bytes-spine}), the observation window, the liveness definition (Section~\ref{sec:residencyworthy}), the cache-object key (Section~\ref{sec:placement-capacity}), and the pooling of the replay budget across replicas (Section~\ref{sec:coldmiss}). We describe each analysis choice alongside the corresponding result.

Studying the open problems above requires workloads with the structure measured here; because the production traces cannot be released, the next section addresses this gap.

\section{AgenticTraceSim: Reproducible Agentic Traces}
\label{sec:agentic-trace-sim}

\label{sec:reproducibility}The production traces contain sensitive serving telemetry and cannot be released without risking client and operational confidentiality. We therefore plan to release \AgenticTraceSim{}, a parametric trace generator that captures the principal distributions and dependencies observed in \Research{} without exposing production events or user content, together with a residency replay (LRU, FNU, and an unbounded-tier MRR variant whose int4 tier never fills) that accepts synthetic or user-supplied call streams.

\AgenticTraceSim{} models six workload components: time-varying session arrivals, session depth (a lognormal mixture), context growth, model-class transitions, per-call reuse, and idle gaps. The components are coupled to preserve the relationships important to residency, including depth with peak context and duration, and \Spine{} reuse with intervening model-class excursions. The generator exposes workload knobs for cache effectiveness, the number of model classes, and the session arrival rate, and the planned release also includes SR and SD population profiles, refitted to those subsets, for sensitivity analysis; these do not model deployment transitions and are not a placement-policy simulator.

We validate both the generated distributions and their principal dependencies. For depth, idle gaps, and per-call reuse, the Kolmogorov--Smirnov distance from the measured \Research{} distributions is at most $0.10$. Context is less closely matched at the per-call level because it is calibrated to reproduce the residency-relevant per-session peak: generated P50/P90/P99 working sets are $21.6/55.9/82.5$~GiB, versus $25.7/59.2/81.6$~GiB measured. The generator also reproduces the measured depth--peak-context and depth--duration rank correlations within $0.05$.

\begin{figure}[t]
\begin{minipage}[c]{0.45\linewidth}
  \centering
  \includegraphics[width=1\linewidth]{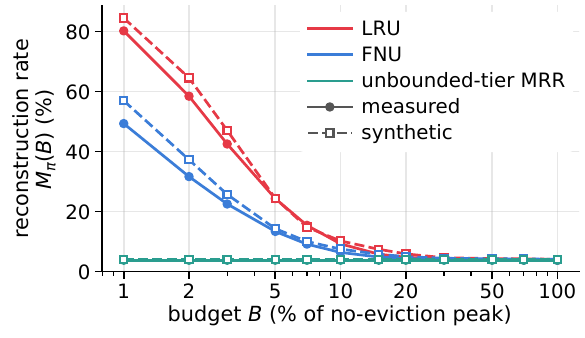}
\end{minipage}\hfill
\begin{minipage}[c]{0.5\linewidth}
\caption{\textbf{\AgenticTraceSim{} preserves downstream residency-policy behavior.} Matched measured and synthetic workloads retain the ordering LRU $\geq$ FNU $\geq$ unbounded-tier MRR at every evaluated budget. This MRR variant uses a separate int4 tier that never fills, unlike the capacity-shared variant in Figure~\ref{fig:coldmiss}. The measured and generated LRU-to-multi-resolution improvements are $2.45\times$ and $2.50\times$, respectively, with overlapping bootstrap intervals.}  \label{fig:gauss_downstream}
\end{minipage}
\vspace{-0.1in}
\end{figure}

Synthetic traces also preserve the downstream residency conclusions. Under the same replay, with budgets as fractions of each stream's own no-eviction peak (not $K_{\mathrm{hr}}^\star$, so LRU rates are not comparable to Section~\ref{sec:coldmiss}'s $62.9\%$), matched real and synthetic workloads retain the reconstruction-rate ordering LRU $\geq$ FNU $\geq$ unbounded-tier MRR at every evaluated budget (Figure~\ref{fig:gauss_downstream}). This unbounded-tier variant of multi-resolution retention demotes displaced KV to a separate int4 tier sized to the no-eviction peak, so the tier never fills and its rate equals the first-touch floor; it is not the capacity-shared variant of Section~\ref{sec:mrr}. The LRU-to-multi-resolution improvement is $2.45\times$ on the matched measured \Research{} sample and $2.50\times$ on the synthetic trace, with overlapping bootstrap intervals. A marginals-only generator fails to reproduce this improvement, showing that the modeled dependencies are necessary for policy evaluation.

The generator and replay produce the synthetic side of every comparison in this section, including the LRU $\geq$ FNU $\geq$ unbounded-tier MRR ordering. The generator reproduces shapes, ratios, and orderings, not absolute fleet magnitudes, which depend on production arrival volume, observation window, and placement topology. The measured side of these comparisons, the capacity-shared MRR replay of Section~\ref{sec:mrr}, and the deployment-local routing and placement results (the generator emits no deployment identifiers) require the production traces; the production values in Sections~\ref{sec:hurdles} and~\ref{sec:capacity} therefore remain measured results rather than generator outputs.

The \AgenticTraceSim{} trace generator and additional appendices offering supplementary material on the trace analysis will be released in the near future.

\section{Related Work}
\label{sec:related}

\paragraph{Agentic workload characterization.} TraceLab characterizes Claude Code and Codex sessions, the GitHub Copilot study measures production coding-agent traffic at scale, and the ReAct study instruments ReAct-style agents on five benchmarks~\cite{zhu2026tracelab,liu2026copilot,yuan2026workload}; Section~\ref{sec:prior-misses} and Table~\ref{tab:vsprior} compare them with this work. KVCache in the Wild characterizes production KV-cache reuse in single- and multi-turn requests at a large cloud provider~\cite{wang2025kvcache}; we focus on multi-call agentic sessions and their deployment placement.

\paragraph{State retention, migration and caching bounds.}
CachedAttention and Pensieve retain multi-turn KV state across memory
tiers~\cite{gao2024cachedattention,yu2025pensieve}, while HCache reconstructs KV
state from stored intermediate activations~\cite{gao2025hcache}. Llumnix
supports cross-instance request and KV-cache migration~\cite{sun2024llumnix}.
These systems provide candidate mechanisms for the capacity and locality
requirements characterized here; we do not compare their implementation-level
latency or transfer costs. Moreover, because FNU is not optimal for
variable-size objects, we treat it only as a clairvoyant reference policy;
flow-based offline bounds provide a more principled optimality benchmark but
are outside our present replay~\cite{berger2018practical}.

\paragraph{LLM serving and KV management.} PagedAttention and vLLM established paged KV management for high-throughput inference~\cite{kwon2023pagedattention}; Sarathi-Serve co-schedules prefill and decode with chunked prefills~\cite{agrawal2024sarathi}, whereas Splitwise, DistServe, and Mooncake run the two phases on separate resources and transfer KV between them~\cite{patel2024splitwise,zhong2024distserve,qin2025mooncake}. Prefix-caching systems improve reuse via structured caching, cache-aware scheduling, and distributed KV transport~\cite{zheng2024sglang,gim2023promptcache,srivatsa2024preble,liu2024cachegen,liu2025lmcache}. These systems provide KV-management mechanisms; we measure the session- and fleet-level demand they must serve in production agentic workloads.

\paragraph{Agent-aware serving and orchestration.} Parrot, Teola, ALTO, Autellix, and Pie optimize serving for whole LLM applications rather than individual requests~\cite{lin2024parrot,tan2024teola,santhanam2024alto,luo2025autellix,gim2025pie}. More recent systems use agent semantics for KV prefetching, pruning, and cache management, and for admission control, program-aware scheduling, and GPU allocation~\cite{pan2025kvflow,li2026intentkv,biswas2026sutradhara,chen2026concur,kang2026thunderagent,wagenlander2026scepsy}. InferCept and Continuum decide whether to keep, discard, or swap a request's KV while it waits on a tool call~\cite{abhyankar2024infercept,li2025continuum}, a per-request counterpart of the idle cap in Section~\ref{sec:residencyworthy}. Model-routing systems choose among models, reasoning modes, or quantized model variants to balance quality, cost, and latency~\cite{ong2025routellm,wang2025semanticrouter,dai2025aragog,taha2026agentcompress}. Our measurements inform these directions by quantifying model-class predictability, idle-gap structure, deployment-conditioned reuse, and the separate reconstruction costs of limited memory and routing.

\paragraph{Context reduction and external memory.} Prompt and context compression reduce the state that must be materialized into KV~\cite{xu2023recomp,pan2024llmlingua2,mu2023gist,kang2026acon}, while agent-memory systems externalize information beyond the active context window~\cite{packer2023memgpt,chhikara2025mem0,rasmussen2025zep,xu2025amem,li2025memos}. These approaches determine what enters the model context; %
we characterize how the resulting KV state is reused, retained, and placed across a production serving fleet; Section~\ref{sec:op-reversible} discusses how the two compose.

\paragraph{KV compression and tiering.} When KV capacity is insufficient, prior work evicts selected tokens~\cite{zhang2023h2o,liu2023scissorhands,xiao2024streamingllm,li2024snapkv,ge2024fastgen}, quantizes retained state~\cite{liu2024kivi,hooper2024kvquant,yue2024wkvquant}, or moves state across memory tiers~\cite{sheng2023flexgen,lee2024infinigen,liu2024cachegen,yao2026vericache}. These techniques define representations and storage tiers for displaced KV; our trace-driven replay asks a complementary workload-level question: how much reconstruction remains under destructive eviction, and how much of it a retained lower-precision copy removes.

\section{Conclusion}
\label{sec:conclusion}

We characterized the KV working set of four production Microsoft M365 workloads spanning three agentic applications and enterprise chat. Agentic sessions repeatedly re-enter LLMs through internal model-tool loops, accumulate context, and reuse much of it across idle intervals; in \Research{}, a P99 session's estimated KV exceeds one H100's HBM, and a predictable \Spine{} class holds most call-context KV bytes. Retaining all of this state is costly, and the resulting memory pressure is not determined by per-session size alone: deep agentic sessions create large individual working sets, while enterprise chat creates substantial fleet demand through concurrency.

Reclaiming state after one idle hour lowers the \Research{} hold-all-live ceiling by $3.6\times$ while raising the replayed reconstruction rate from $4.07\%$ to $4.53\%$. Beyond idle reclamation, the busiest-hour residency replay identifies two distinct sources of reconstruction. %
Under a pooled memory budget equal to $10\%$ of the busiest-hour current-context peak, even clairvoyant destructive eviction by Belady's furthest-next-use rule (FNU) reconstructs context for $31.1\%$ of \Research{} calls, while int4 MRR, which keeps evicted KV as a recoverable copy, reduces this rate to $13.6\%$. Separately, even with unlimited memory, modeling KV as deployment-local raises the first-touch floor from $5.4\%$ to $20.2\%$. Efficient agentic serving therefore requires both memory mechanisms that preserve recoverable KV under capacity pressure and placement mechanisms that maintain or move reusable state across deployments.

Because the production traces contain sensitive enterprise telemetry, we plan to release \AgenticTraceSim{} and a residency replay for synthetic or user-supplied call streams, for reproducible counterfactual studies. The generator preserves the workload dependencies needed to reproduce the reconstruction-rate ordering of LRU, FNU, and unbounded-tier MRR without exposing production events. Together, the measurements and the planned artifacts support the study of fleet-scale KV residency and routing policies that account for session working-set size, expected reuse, idle duration and deployment placement.

\clearpage
\bibliographystyle{plain}
\bibliography{references}

\end{document}